\documentstyle[aps]{revtex}

\begin{document}
\draft
\preprint{}
\title{    
Effect of the Fermi surface destruction on transport properties in 
underdoped cuprates}
\author{
Jian-Xin Li$^{*}$}
\address{
Institute of Physics, Academia Sinica, Taipei 11529, Taiwan}
\author{W.C. Wu}
\address{Institute of Physics, Academia Sinica, Taipei 11529, Taiwan\\
and Department of Physics, National Taiwan Normal University,
Taipei 11718, Taiwan}
\author{
T.K. Lee}
\address{
Institute of Physics, Academia Sinica, Taipei 11529, Taiwan \\
and National Center for Theoretical Sciences, P.O.Box 2-131, Hsinchu, Taiwan}

\maketitle

\begin{abstract}
Motivated by recent experimental measurements on the Fermi surface(FS) destruction 
in underdoped high-$T_{c}$ cuprates, we examine its effect on the transport
properties based on the Boltzmann equation approach. The effect is modeled
by simply taking the density of states for electrons in the gapped regions
to be zero. Within the nearly antiferromagnetic Fermi liquid model, we
calculate the temperature dependences of the dc resistivity, the inverse
Hall angle and the Hall coefficient. It is 
shown that the effect of the FS destruction on transport properties is 
sensitive to the existance and the range of the flat band near $(0,\pm\pi)$
in the dispersion of electrons, and the anistropy of the relaxation rate
along the Fermi surface.
We find that the experimental data are better described by the cold spot
model, i.e., the transports are determined mainly by the contribution of
the electrons near the Brillouin-zone diagonals.

\end{abstract}


\section{INTRODUCTION}
Normal state transports in high-$T_{c}$ cuprates continue to be a
challenging subject. It is known for a long time that the resistivity
$\rho(T)$ in the normal state shows a linear temperature behavior down to
the superconducting transition temperature $T_{c}$ and meanwhile the inverse 
Hall angle $\cot \theta_{H}(T)$ and the Hall coefficient $R_{H}(T)$ have 
$T^{2}$ and $T^{-1}$ temperature dependences, respectively. 
However, the situation is different in underdoped systems, in which both 
$\rho(T)$ and $R_{H}(T)$ deviate from their high-$T$ behaviors 
below certain temperatures higher than $T_{c}$~\cite{buc,wuy}.

Various models have been proposed to accout for the temperature behaviors 
of the resistivity, the inverse Hall angle as well as the Hall 
coefficient~\cite{sto}.
Among them, of accumulating interest is the models based on the so-called  
hot spots and/or cold spots~\cite{rise,pine,iof,zhe}, which
refer to small regions on the Fermi surface(FS) where the 
electron lifetime is unusually short or long, respectively. 
Fundamentally, in this kind of models, the anomalous temperature 
dependences of transport coefficients are ascribed to the anistropy of 
scatterings on different momentum regions. Some successes have been achieved
based on these models. However, there are relatively few studies of the 
transport properties in underdoped cuprates. One of the striking features in
underdoped high-$T_{c}$ cuprates is that there is a normal state gap
(pseudogap) as measured by various experiments~\cite{ran}. Recent
angle-resolved photoemission(ARPES) experiment~\cite{nor} further indicates
that the pseudogap opens up at different momentum points at different
temperatures, consequently it leads to a FS composed of disconnected arcs
in the pseudogap state. Because the electron lifetime varies over the Fermi
surface as assumed in the hot spot and/or cold spot models and also as
suggested by ARPES experiments~\cite{shen,gof}, it is
expected that the losing of some parts of FS will affect its
temperature behavior. It is our aim in this paper to study the
effects of the destruction of the FS on the transport properties in the 
pseudogap state. Our main results are: (1) Based on the standard Boltzmann 
transport theory, we demonstrate that the transport properties in the 
pseudogap state are well described by the cold spot model--- the main 
contribution to transports comes from the cold spots and the hot spots
contribute little, which is consistent with the recent studies
on the transports in the normal state~\cite{sto,iof}. (2) For the realistic
calculations using the nearly antiferromagnetic Fermi liquid (NAFL) 
interaction form~\cite{sto,pine,mon}, we find that the variation of
the Fermi velocity along the FS is an essential ingredient for the 
justification of the applicability of the cold spot model. (3) The 
bandstructure which has an extended flat band near $(0,\pi)$ gives a
good account for the experimental observations. (4) By reducing the
dispersion for optimally doped high-$T_{c}$ cuprates by a factor of 3,
we can fit our result for the resistivity with experiments quantitatively.
Moreover, using the same parameters, we find that the calculated temperature
dependence of the inverse Hall angle is also consistent with the 
experimental data~\cite{wuy}. As for the Hall coefficient, we get a weaker 
temperature dependence than experiments, however its crossover behavior
from the normal to the pseudogap state is in agreement with 
experiments qualitatively.

The paper is organized as follows. In Section II, we discuss the effect
of the variation of the Fermi velocity along the FS on the resistivity in the 
pseudogap state by comparing two kinds of tight-binding bandstructures 
which differ in the flatness of the dispersions
near $(0, \pm\pi)$ points. In Section III, we present fits to experimental
data of the resistivity and the inverse Hall angle and discuss qualitatively
the crossover behavior of the Hall coefficient from the normal to the
pseudogap state. Section IV contains a brief discussion and a conclusion.

\section{EFFECT OF THE VARIATION OF FERMI VELOCITY ON RESISTIVITY}

Currently, the most commonly used band structure for quasiparticles in
high-$T_{c}$ cuprates is the
two-dimensional tight-binding model including the neast- and next-neast-
neighbour hopping term which is written as,
\begin{equation}
\varepsilon_{k}=-2t(\cos k_{x}+\cos k_{y})
-4t^{'}\cos k_{x}\cos k_{y}-\mu,
\end{equation}
where, $t=0.25$eV, $t^{'}/t=-0.45$ and $\mu$ is the chemical potential 
which is determined by hole concentration. As will be discussed below, we
find that this dispersion fails to account for the temperature
dependence of the resistivity
in the pseudogap state as far as our model is concerned. Thus,
another bandstructure is also considered, which is obtained
by a tight-binding fit to ARPES energy dispersion by 
Norman {\it et al.}~\cite{nor}. It reads,
\begin{eqnarray}
\varepsilon_{k}&=&t_{0}+t_{1}(\cos k_{x}+\cos k_{y})+t_{2}\cos k_{x}\cos k_{y}
+t_{3}(\cos 2k_{x}+\cos 2k_{y}) \\ \nonumber
 & &+t_{4}(\cos 2k_{x} \cos k_{y}+\cos 2k_{y} \cos k_{x})
+t_{5}\cos 2k_{x}\cos 2k_{y}-\mu,
\end{eqnarray}
with real space hopping matrix elements (in eV) [$t_{0},...,t_{5}$]= 
[0.1305,-0.2976,0.1636,-0.026,-0.0559,0.051]. Experimentally, 
the pseudogap opens at different temperatures for different hole doping 
concentrations. However, since the opening temperature $T^{*}$ is chosen by  
hand in our model, this change can be naturally realized. Thus, we will
not consider the effect of different doping levels and fix the hole
concentration $n=0.1$. The FS's for these two dispersions corresponding to 
$n=0.1$ are shown in Fig.1. The main difference
between them is that the energy band Eq.(2) is flatter near the crossing 
of the FS and the Brillouin zone boundary. This difference can be seen more
clearly from their dispersions plotted in Fig.2. As shown, a very flat band
exists near the ${\rm M}$ point along the direction of ${\rm \Gamma}$ to
${\rm M}$ for Eq.(2). Consequently, the Fermi velocity at
$k$-point $A$ for the dispersion (2) is nearly 2.5 times smaller than at 
$k$-point $B$($A$ and $B$ are indicated in Fig.1), while it varies slightly
for the dispersion (1), as one can see from the inset of Fig.2. In the
following, we will see that this slight difference affects the temperature
behavior of the transport coefficients in the pseudogap
state qualitatively. It is worthy to point out that both dispersions are 
obtained by fitting to the ARPES experiments on the optimally doped materials.
So, applying them to underdoped systems by just adjusting the chemical 
potential is a rigid band approximation. Because no detailed bandstructure
for underdoped materials is available for us now, we will use this 
approximation in this section for a qualitative discussion on the 
sensitivity of the resistivity in the pseudogap state with respect to
the variation of the Fermi velocity along the FS. In section IV, we will 
demonstrate that this assumption fails to fit to the experimental data 
quantitatively and the best fit to experiments is the dispersion (2)
reduced by a factor of 3.

Though there are many studies for the origin of this pseudogap~\cite{lee}, 
no consensus seems to have been achieved. The ARPES experiments
show that the gap has a $d_{x^{2}-y^{2}}$ symmetry and it first appears
near $(0,\pm \pi)$ and $(\pm \pi,0)$ points, the gapped regions spread
laterally on cooling the samples~\cite{nor}.  In the presence of gap,
the transfer rates of electrons into and out of these regions as well as the 
excitations of electrons in these regions will drop rapidly. For simplicity, 
here we assume that the states in the gapped region( shown schematically 
in Fig.1 as the regions closed by four bold line
semi-circles) is unavailable for electrons. These regions first appear at 
the opening temperature  of the pseudogap
and will extend gradually as temperature decreases. 
Because only several ARPES data on the destruction of the FS are available, 
we can not deduce a precise form of its variation as a function of 
temperature and will assume that the $T$-dependences of the radius of 
the gapped regions will be $R(T)\propto (T^{*}-T)$,
$R(T)\propto (T^{*}-T)^{1/2}$ and $R(T)\propto \tanh 2\sqrt {(T^{*}/T)-1}$ 
($T^{*}$ the opening temperature of the pseudogap) and its maximum value
at the superconducting transition temperature $T_{c}$ be $R_{max}=0.3\pi$
(the case of $R_{max}=0.25\pi$ is sometimes also included for comparison).
We will choose $T^{*}=150$K and $T_{c}=$64K to fit to the experimental
data on YBa$_{2}$Cu$_{3}$O$_{6+x}$.  Our model is
reminiscent of a recent proposal by Furukawa, Rice and Salmhofer~\cite{fur}. 
Based on the one-loop renormalation group investigation, they demonstrated
that the FS with saddle points $(\pi,0)$ and $(0,\pi)$ can be truncated by 
the formation of an insulting condensate due to the umklapp scattering as
the electron density increases, while the remaining FS is still metallic. 
It also bears a close similarity to the bosonic preformed pairs model
by Geshkenbein, Ioffe and Larkin~\cite{ges}, in which the fermions lying 
inside the disks shown in Fig.1 are assumed to be paired into dispersionless
bosons, and the interaction of transferring electrons from the disks to other
parts of the FS is weak so that the bosons are in fact localized.

In order to proceed the detail calculations, a form of the effective
interaction between electrons and spin fluctuations is required. We
note that, the anisotropy of scatterings on the Fermi line can be naturally
realized in the nearly antiferromagnetic Fermi liquids(NAFL) 
model~\cite{sto,pine,mon}, where the spin fluctuations strongly peak at the
AF wave vector $(\pi,\pi)$. So, we will adopt this model interaction
which reads~\cite{sto,pine,mon},
\begin{equation}
\chi({\bf q},\omega)=\sum_{i} {1 \over {\omega_{q_{i}}-i\omega}}
\end{equation}
where $\omega_{q_{i}}=T^{c}+\alpha T+\omega_{D}\psi_{q_{i}}$, 
$\psi_{q_{i}}=2+\cos(q_{x}+ \delta  Q_{i})+\cos(q_{y})$(or 
$2+\cos(q_{x})+\cos(q_{y}+\delta Q_{i})$), $T^{c}, \alpha,$ and 
$\omega_{D}$ are temperature-independent parameters. The sum
over $i$ runs over the incommensurate wavevector 
$\delta Q_{i}=\pm 0.12\pi$, which has been shown to exist in
YBa$_{2}$Cu$_{3}$O$_{6.6}$~\cite{dai}, recently. In our discussion, the
results with commensuration $\delta Q_{i}= 0$ are also
included for comparison.

We assume that a weak magnetic field is applied perpendicular to the 
CuO$_{2}$ plane and an electrical field is along the $x$ direction, i.e.,
${\bf E}=E{\bf e_{x}}$ and ${\bf B}=B{\bf e}_{z}$. Within the
conventional relaxation-time approximation, the longitudinal and Hall
conductivities can be calculated according to,
\begin{equation}
\sigma_{xx}=-2e^{2}\sum_{k}[{\bf v}(k)\cdot {\bf e_{x}}]^{2} \tau_{k}
[{{\partial f(\varepsilon_{k})} \over \partial \varepsilon_{k}} ],
\end{equation}
\begin{equation}
\sigma_{xy}=-2e^{3}\sum_{k}[{\bf v}(k)\cdot {\bf e_{x}}\tau (k)] {\bf v}(k)
\times {\bf B}\cdot \nabla [{\bf v}(k)\cdot {\bf e_{y}}\tau (k)]
[{{\partial f(\varepsilon_{k})} \over \partial \varepsilon_{k}}],
\end{equation}
where, ${\bf v}(k)=\nabla _{\bf k}\varepsilon_{k}$ is the group velocity and 
$\tau (k)$ the relaxation time. 

Following Stojkovic and Pines~\cite{sto}, we will approximate the relaxation
rates by the electron lifetime. To second-order in the interaction
constant $g$, it reads,
\begin{equation}
{1\over \tau (k)}=2g^{2}\sum_{k'} {\rm Im}\chi(k-k',\varepsilon_{k'}-
\varepsilon_{k})[n(\varepsilon_{k'}-\varepsilon_{k})+f(\varepsilon_{k'})],
\end{equation}
where $n(\varepsilon)$ and $f(\varepsilon)$ are the Bose and Fermi 
distribution functions, respectively.

We solve Eqs.(4), (5) and (6) numerically by dividing the 
Brillouin zone into 200$\times$200 lattices.
The parameters for the spin-fluctuation spectrum are choosen as
$T^{c}=0$, $\alpha=2.0$ and $\omega_{D}=77$meV.  Another set of
parameters, namely, $T^{c}=0$, $\alpha=2.0$ and $\omega_{D}=147$meV~\cite{mon}
is also used, no qualitative change has been found. The interaction
constant is taken to be $g=0.64 eV$ as used before~\cite{mon}.
In the pseudogap
state, the sum over $k(k')$ in Eqs.(4), (5) and (6) will exclude those
regions where the gap is formed.

The temperature dependences of the relaxation rates $\tau_{k}$ for both 
dispersions are presented in Fig.3. Due to the opening of the pseudogap, 
a decrease is observed for all cases at low temperatures.  This result
is expected because we have taken the density of states in the
gapped region to be zero after the temperature is lower than the opening
temperature $T^{*}=150$K.  We note that the gapped regions
first appear at $(0,\pm\pi)$ points and the energy difference between the
chemical potential and that at $(0,\pm\pi)$ is 1950 K and 650 K for the 
dispersions (1) and (2), respectively, at the same doping concentration
$n=0.1$. Because transports involve the scatterings of electrons situating
about several K$T$ around the FS, the destruction of the FS affects the
transport properties only when the difference between the energies at the
FS and at the gapped regions is comparable with K$T$. As a result, the
decrease in $1/\tau_{k}$ starts at different
temperatures for the dispersions (1) and (2). It starts at nearly $T^{*}$
for the dispersion (1)(shown in (c) and (d)) and at lower temperature for
the dispersion (2)(shown in (a) and (b)), especially it will depends on
the spreading rate of the gapped regions for the dispersion (2). Comparing
$1/\tau_{k}$ at different $k$-points $A$ and $B$, one finds that the
relaxation rate is strongly anistropy along the FS, it is larger near the hot
spots such as the $k$-point $A$ and smaller near the cold spots such as
$k$-point $B$. This is due to the anistropy of the interaction form Eq.(3).
We would like to emphasize that there is no appreciable difference in
the ratios of the $1/\tau_{k}$ at hot spots to at cold spots calculated using
the dispersions (1) and (2), namely it is about 4 for the dispersion (1) and
7 for the dispersion (2).

Now, we turn to the discussion of the dc resistivity. Before proceeding with
a detail analysis, one may speculate that the resistivity
will decrease once the temperature is below $T^{*}$(or the temperature when 
$1/\tau_{k}$ starts to decrease for the dispersion (2)) as inferred from the 
behavior of $1/\tau_{k}$, according to the well-known Drude formular for 
resistivity $\rho=m^{*}/(n_{e}e^{2}\tau)$ (here $\tau$ is an
effective relaxation rate, $m^{*}$ the effective mass and $n_{e}$ the number
of electrons). However, the numerical result for $\rho (T)=1/\sigma_{xx}$
calculated using Eq.(4) turns out to be not so trivial, since the quantity
$m^{*}/n_{e}$ will change after parts of the FS is destroyed due to the
formation of the pseudogap. As shown in Fig.4(a), the dc resistivity
calculated using the dispersion (1) goes up instead of going down after
the gap opens for all cases of $R_{max}$ and $R(T)$, which contradicts 
the experimental observation~\cite{buc},
while those calculated using the dispersion (2) (as shown in Fig.4(b)) 
show a decrease below $T^{*}$ though a small rise can also be observed 
below about 100 K and 80 K for the cases of $R_{max}=0.3\pi$,  
$R(T)\propto (T^{*}-T)$, and $R_{max}=0.25\pi$, $R(T)\propto \tanh 2\sqrt
{(T^{*}/T)-1}$, respectively. Since the relaxation rate at the cold spots
shows decrease once entering into the pseudogap state, the contribution to
the longitudinal conductivity from the cold spots $\sigma_{xx}^{(c)}$ will
increase. On the other hand, the density of states near the hot spots will
lose because
of the opening of pseudogap and it gives rise to a decrease in the
conductivity coming from the hot spots $\sigma_{xx}^{(h)}$. Therefore,
whether the resistivity rises or drops in the pseudogap state depends
on the competition of the increment in $\sigma_{xx}^{(c)}$ and the
decrease in $\sigma_{xx}^{(h)}$. For the cold spot model in which the
relaxation rate at the hot spots is assumed to be unusually larger than
at the cold spots~\cite{sto,iof}, $\sigma_{xx}^{(h)}$ will be
short-circuited and the conductivity $\sigma_{xx}$ is determined by
$\sigma_{xx}^{(c)}$. So, $\sigma_{xx}$ will increase and in turn the
resistivity $\rho(T)=1/\sigma_{xx}$ will decrease. To the contrary,
for the hot spot model~\cite{zhe} the contribution from the hot spots is
comparable with or even larger than that 
from the cold spots, then the decrease in $\sigma_{xx}^{(h)}$ will surpass
the increase in $\sigma_{xx}^{(c)}$ and it leads to a drop in the
conductivity and a rise in the resistivity. Thus, in order to account for
the temperature behavior in resistivity observed in the pseudogap state,
the resistivity should be dominated by the contribution from the cold spots
and that from the hot spots be negligible. 
Now, we return to our realistic calculation using the interaction form
Eq.(3). From Fig.3, one finds that the ratio of the relaxation rate at the
hot spots to that at the cold spots is about 4 ( a band calculation gives
the same ratio, see table II in Ref.~\cite{zhe}) and 7 for the dispersions
(1) and (2), respectively. Thus, no overwhelming contribution from the
cold spots can be expected just from this ratio. In this case, the
kinematical factor (Fermi velocity $v_{F}$) should be considered, since the
transport coefficients involve a $k$-sum over $\tau_{k}$ weighted
by $v_{F}^{2}$. As noted above, for the dispersion (2) the ratio of the
Fermi velocity at the cold spots (near $k$-point $B$) is 2.5 times larger
than at the cold spots (near $k$-point $A$). This, along with the ratio
7 for the relaxation rate, makes $\sigma_{xx}^{(c)}$ at the $k$-point $A$
be 44 times larger than $\sigma_{xx}^{(h)}$ at the $k$-point $B$ and
justifies the applicability of the cold spot model. However, for the
dispersion (1) the Fermi velocity at the cold spots is nearly 1.15 times
smaller than at the cold spots, consequently $\sigma_{xx}^{(c)}$ is only
3 times larger than $\sigma_{xx}^{(h)}$. Thus, the losing in
$\sigma_{xx}^{(h)}$ due to the opening of the pseudogap will exceed the
increase in $\sigma_{xx}^{(c)}$ arising from the enhancement in
$\tau_{k}^{(c)}$ and eventually the resistivity will increase.

From the above discussion, one can see that the crossover behavior of the
resistivity is better described by the cold spot rather than the hot spot
model, which is consistent with the recent studies on the transport
properties in the normal state~\cite{sto,iof}. In a realistic calculation,
we find that the variation of the Fermi velocity along the FS plays an
important role in the determination of the cold spot or hot spot model.
In terms of the ARPES experiments~\cite{shen,gof}, an extended 
van Hove singularity (flat band) exists near $(0,\pm \pi)$ --- around the
hot spots, it will
lead to a lower Fermi velocity around the hot spot region and justifies the 
applicability of the cold spot model. However, the energy dispersion of
Eq.(1) is not flat enough and meanwhile the flat band is far away from
the FS. Consequently, it has larger Fermi velocity near the hot spots as
shown in the inset of Fig.2. So, as far as our model is concerned, the
dispersion (1) is inadequate for the description of the transport properties
in the underdoped cuprates though it was used mostly before.

\section{FITTING TO EXPERIMENTAL DATA}

Although the agreement between the model calculation of the resistivity 
using the dispersion (2) and experiments is reasonable in view of its
crossover behavior from the normal to the pseudogap state, there are
two discrepancies when fitting it to experiments quantitatively. One is
a non-linear temperature resistivity appears at high temperatures and
the other is that the resistivity ceases to decrease and even has a
slight rise with further decreasing temperature below about 100 K
which is higher than $T_{c}$ as can be seen in Fig.4(b).
To resolve these discrepancies, we note that the
bandstructure (2) is obtained from a fit to the photoemission 
experimental data of Ba$_{2}$Sr$_{2}$CaCu$_{2}$O$_{8}$ with hole doping
0.17, i.e., an optimally doped cuprate, so applying it to the underdoped
regime by just adjusting its chemical potential is a rigid band assumption.
In fact, the bandstructure will change as doping varies. An important
fearure is that the band width will become narrow, i.e., the quasiparticles 
will becomes heavy as doping decreases. Of course, this renormalization of 
mass is anisotropic in momentum space, it is larger near the FS and becomes 
more and more less away from it. The detailed treatment 
of this renormalization requires a complicated calculation and goes beyond 
our scope here, thus we simply take $\varepsilon_{k} \rightarrow 
\varepsilon_{k}/3.0$. This amounts to reducing the energy difference 
between the chemical potential and the flat band for hole doping $n=0.1$ 
from 54 meV to 18 meV, which is consistent with ARPES experimental data 19 
meV for the underdoped YBa$_{2}$Cu$_{4}$O$_{8}$~\cite{gof}. The reduction of
the whole
energy band by a factor of 3 is justified approximately by the fact that
just the electrons near the FS contribute to transports and those 
far away from it have in fact no effect. Before making a 
quantitative comparison with experiments, we note that Wuyts 
{\it et al.}~\cite{wuy} have developed an universal analyzing method for 
transport data in underdoped high-$T_{c}$ superconductors. They demonstrated 
that the transport data on YBa$_{2}$Cu$_{3}$O$_{x}$ can be scaled onto an 
universal curve using one scaling parameter $T_{0}$ which has 
$0.8T_{0} \approx T^{*}$. We will adopt this method in the following analysis.

The results for the dc resistivity $\rho (T)=1/\sigma_{xx}$,
calculated using the dispersion (2) reduced by a factor of 3 with $\mu=0.02t$
($n$=0.1) and using the dispersion (2) but with $\mu=-0.016t$( the
dashed-dotted line), are shown in
Fig.5, where the residual resistence is taken to be $\rho_{0}=0.162 
\rho(T_{0})$, the same value as used before~\cite{wuy}, $T^{*}=0.72T_{0}$ 
with $T^{*}=150$K. The result represented by the dashed-dotted line has
the same energy difference 18 meV to that calculated using the reduction
of the bandwidth. The hollow squares indicate the experimental data of
Ref.~\cite{wuy}. An important effect of the reduction of the energy band
is that the flat region becomes large, so the slight rise in the resistivity
below about 100 K observed in Fig.4(b) is removed and a continue decrease
is obtained which fits the experimental data well. On the other hand, though
having the same energy difference, the result by adjusting the chemical
potential (dashed-dotted line) still shows a rise below 100 K. It implies
that it is the range of the flat band instead of the difference between
the chemical potential and the flat band that has something to do with
the low temperature rise in resistivity. This may be understandable
from the reason causing this rise. As the gapped regions spread with
decreasing temperature, more and more parts of the FS are destroyed. Since
the Fermi velocity increases when the wavevector moves from the $k$-point
$A$ to $B$, the ratio of the Fermi velocity at the cold
spots to that at the crossing of the FS and the edge of the gapped regions
will decrease. Thus, the contribution to the conductivity from the hot
regions will grow gradually and the resistivity will cease to decrease or
even rise in low temperatures. If we reduce the width of the energy band,
then the range of the flat band and consequently the range of the low Fermi
velocity will grow. It enables the contribution from the hot regions 
to be negligible. One can also see from the figure that a linear in $T$
dependence is well reproduced in the normal state, although its slope is
somewhat larger than experimental
data. The results for different $T$-dependences of the radius of gapped 
regions $R(T)$ are shown in the inset of Fig.4. We find that the best fit to    
the experimental data is $R(T)\propto (T^{*}-T)$ though the difference between 
$R(T)\propto (T^{*}-T)^{1/2}$ and $R(T)\propto (T^{*}-T)$ is minor.
The comparison of the resistivities calculated for the commensurate and
incommensurate cases is also shown in Fig.5, in which the dashed line
represents the result for the commensurate case. The qualitative difference
between them is that the slop is smaller for the commensurate case
and make the fit become bad. As noted above, the temperature dependence of
the resistivity is determined by the weight of the contributions from the
cold spots to that from the hot spots. The incommensurate wave vectors will
make the hot spots be shifted away from the original ones (move to the
cold spots for $-\delta {\bf Q}$ and to the boundary of the Brillouin zone
for $\delta {\bf Q}$) and thus lead to a change of the weight of the
contribution from the cold spots to that from the hot spots. It is this
change that gives rise to different slops in the
temperature behavior of the resistivity. 

Using the same parameters, we have calculated the $T$-dependences of the
inverse Hall angle $\cot \theta_{H}(T)=\sigma_{xx}/\sigma_{xy}$ and
the Hall coefficient $R_{H}=\sigma_{xy}/B\sigma_{xx}\sigma_{yy}$.
The result for the inverse Hall angle is presented in Fig.6. 
The fit to experimental data above $T^{*}$ is good below about 300K.
Below 150 K, a slight deviation from the $T^{2}$ appears, especially,
the $\cot \theta_{H} (T)$ curve changes from convexity to concavity with
nearly the same reflection point as the experimental data for the
incommensuration case. Comparing the results for different $T$-dependences of
the gapped regions, one can find that there is no significant difference
as shown by the solid and dotted lines. The results for the commensurate
case (dashed line) and calculated using the dispersion (2) with
$\mu=-0.016t$(dashed-dotted line) all show disagreement with the
experimental data at high temperatures and also exhibit a deviation from
the high-$T$ behavior at much low temperature compared with the experiment.

As for the Hall coefficient shown in Fig.7, we obtain a weaker temperature
dependence than that seen experimentally~\cite{buc,wuy}. The similar result
has been
reported by Stojkovic and Pines based on the NAFL model~\cite{pine}. However,
the trend is very similar to the experimental data and allows us to
compare its crossover behavior with the experiment qualitatively.
A striking feature for the incommensuration cases which are represented by
the solid and dotted lines is that the
Hall coefficient decreases rapidly with decreasing temperature at low
temperatures and causes a peak occurring slightly below the opening
temperature $T^{*}$. This is qualitatively consistent with
experiments~\cite{buc,wuy}.
We note that the result with commensuration (dashed line) shows an even
weaker $T$ dependence than the case of incommensuration. Moreover, the
results calculated using the dispersion (2) with $\mu=-0.016t$(
dashed-dotted line) displays a contrary temperature behavior, i.e., it
decreases with temperature. This strong discrepancy is related to the
discrepancy in the resistivity discussed above. Because
$R_{H}\propto 1/\sigma_{xx}^{2}$, any slight deviation from the
$T$-linearity in $\rho(T)$ will be amplified and leads to a worse result
for the Hall coefficient. From the same arguement as that for resistivity,
we know that the main contribution to the transverse conductivity
$\sigma_{xy}$ comes from the cold spots and the hot spots contribute
little~\cite{sto}. So, $\sigma_{xy}$ will have the same trend as that for
the longitudinal conductivity $\sigma_{xx}$, and their
effect will cancel and lead to a minor variation in the temperature
dependence of the inverse Hall angle. On the other hand, the depression
will reflect in the Hall coefficient since we have
$R_{H}=\sigma_{xy}/B\sigma_{xx}\sigma_{yy}$.

This agreement with experiments using the dispersion (2) reduced by a factor
of 3 raises a question: whether the dispersion (1) also
works after the same reduction. We have done it and the result turns out to
be bad. The reason is that there is another vH singularity in the
dispersion (1) except that at $(0,\pi)$, it exists at $(0,0)$ point.
That part of the FS near the diagonal direction will approach to this
one when the energy band is reduced and increase the density of states at
the cold spots, eventually this will change the weight of cold spots to
hot spots drastically. 

\section{DISCUSSION AND CONCLUSION}

In summary, we have investigated the effect of the FS destruction on
transport properties in the pseudogap state of underdoped high-$T_{c}$
cuprates based on the standard Boltzmann theory. Using a simple assumption
of taking the density of states of the gapped regions to be zero, we
calculate the temperature dependences of the longitudial resistivity, the
Hall angle and the Hall coefficient. The results indicate that the
temperature dependence of the transport coefficients is strongly sensitive
to the existance and the range of the flat band near $(0,\pm \pi)$, and the
anistropy of the scattering rates for electrons along the Fermi surface.
We find that the temperature dependences of the transport coefficients
in the pseudogap state are better described by the cold spot model, i.e.,
they are determined by the contribution from the cold spots while the hot
spots contribute little. We can semi-quantitatively explain the temperature
dependences of both the resistivity and the Hall angle, as well as
qualitatively explain the crossover behavior of the Hall coefficient from
the normal state to the underdoped state. However, the calculated Hall
coefficient in the normal state shows a weaker temperature dependence
than that observed by experiments.

It is worthwhile to point out that in NAFL model
the different magnetic properties in underdoped systems are ascribed to
distinct scaling regimes of the spin-fluctuation spectrum Eq.(2)~\cite{bar}.
From this point of view, the anomalous transport properties may arise from
the different interaction form which is related to the opening of the
"spin pseudogap" as deduced from the NMR and neutron scattering
experiments~\cite{tak}, though there is no detailed calculations about it
now. Here, we focus on the effect of the Fermi surface topology and
consider the interaction form in the underdoped regime to be the same as
that in the optimally doped regime. What is the relation
between the two proposals and also if any other interaction form which gives
a varying electron lifetime on the FS such as Eq.(2) can give the 
same results presented here deserve further investigations.

\section*{ ACKNOWLEDGMENTS}
We would like to thank M.R.Norman for discussion and for pointing out
the dispersion Eq.(2). We acknowledge the support from NSC of Taiwan under
Grants No.88-2112-M-001-004 and No.88-2112-M-003-004.
JXL is support in part by the National Nature Science Foundation of China.

$^{*}$ On leave from 
Department of Physics, Nanjing University, Nanjing 210093, People's Republic
of China

\newpage
\section*{FIGURE CAPTIONS}
\vspace{0.5cm}
Fig.1  Fermi surfaces for the dispersions (1) (dashed line) and (2) 
(thin solid line) with hole doping $n=0.1$. The thick solid lines enclose
the "disk" regions where a pseudogap is suggested by experiments. The
density of states of electrons in that regions will be assumed to be 
zero in our model calculations.

\vspace{0.5cm}
Fig.2  Energy dispersions for Eqs.(1) and (2) described in the text along 
the $\Gamma=(0,0)$ --- ${\rm M}=(0,\pi)$ --- ${\rm Y}=(\pi,\pi)$ 
direction.
Note the very flat band existing near ${\rm M}$ for Eq.(2). The inset shows
the Fermi velocities along Fermi surface for the dispersions (1) (dotted 
line) and (2) (solid line). The $k$-point $A', B$ and $A$ corresponds to
what indicated in Fig.1.

\vspace{0.5cm}
Fig.3  Relaxation rates as a function of temperature at different $k$ 
points along the Fermi surface. (a) and (b) are the results calculated
using Eq.(1), (c) and (d) are those using Eq.(2). The $k$-point symbols
($A$ and $B$) correspond to what indicated in Fig.1. The maximum value 
of the radius of the gapped region is $R_{max}=0.25\pi$(see text) and their 
$T$-dependences are $R(T) \propto (T^{*}-T)$(solid lines), $(T^{*}-T)^{1/2}$
(dashed lines) and $\tanh(2\sqrt{T^{*}/T-1.0})$(dotted lines).
The results are for 
the commensurate magnetic interaction, those for the incommensuration
case are qualitatively similar to the results shown here except for a larger
values.

\vspace{0.5cm}
Fig.4  Sensitivity of the resistivity with respect to the dispersions
(1) [(a)] and (2) [(b)]. The solid line indicates the result with 
a $T$-dependence of the radius of the gapped region $R(T) \propto 
\tanh(2\sqrt{T^{*}/T-1.0})$, the dashed line with $R(T) \propto
(T^{*}-T)$, and both correspond to the maximum value $R_{max}=0.25\pi$. 
The dotted line corresponds to the case of $R_{max}=0.3\pi$
and $R(T) \propto (T^{*}-T)$. For comparison, we  
also show the result for the commensuration case as indicated in the 
figure.

\vspace{0.5cm}
Fig.5  Scaled resistivity versus scaled temperature 
with the maximum value of the radius of the gapped region $R_{max}=0.3\pi$. 
The solid, dashed, dotted lines and those in the inset are the results 
calculated using the dispersion (2) reduced by a factor of 3. The 
dashed-dotted line is the result calculated using the dispersion (2) with 
the chemical potential $\mu=-0.016t$ (see text). Solid line: the
$T$-dependence of the radius of the gapped region $R(T)\propto (T^{*}-T)$, the
incommensuration $\delta Q=0.12\pi$. Dashed line: $R(T)\propto (T^{*}-T)$,
$\delta Q=0$. Dotted line: $R(T)\propto (T^{*}-T)^{1/2}$, $\delta Q=0.12\pi$.
Dashed-Dotted line: $R(T)\propto (T^{*}-T)$, $\delta Q=0.12\pi$.
The open squares, both in the main panel and in the inset, are
experimental data from Ref.~\cite{wuy}.
Inset shows the sensitivity of the resistivity with respect to the
size and temperature dependence of the gapped region. Solid line: the 
same parameters with the solid line in the main panel. Dashed line:
$R(T)\propto (T^{*}-T)$, $R_{max}=0.25\pi$. Dotted line: $R(T)\propto 
\tanh(2\sqrt{T^{*}/T-1.0})$, $R_{max}=0.25\pi$.

\vspace{0.5cm}
Fig.6. Scaled inverse Hall angle versus scaled temperature with 
the maximum value of the radius of the gapped region $R_{max}=0.3\pi$. 
The solid, dashed and dotted lines are the results calculated using the 
dispersion (2) reduced by a factor of 3. The dashed-dotted line is the 
result calculated using the dispersion (2) with the chemical potential 
$\mu=-0.016t$ (see text). 
Solid line: the $T$-dependence of the radius of the 
gapped region $R(T)\propto (T^{*}-T)$, the incommensuration $\delta 
Q=0.12\pi$. Dashed line: $R(T)\propto (T^{*}-T)$, $\delta Q=0$.
Dotted line: $R(T)\propto (T^{*}-T)^{1/2}$,  $\delta Q=0.12\pi$.
Dashed-Dotted line: $(T^{*}-T)$, $\delta Q=0.12\pi$.
The open squares are experimental data from Ref.~\cite{wuy}.

\vspace{0.5cm}
Fig.7. Temperature dependence of the Hall coefficient with
the maximum value of the radius of the gapped region $R_{max}=0.3\pi$. 
The solid, dashed and dotted lines are the results calculated using the 
dispersion (2) reduced by a factor of 3. The dashed-dotted line is the 
result calculated using the dispersion (2) with the chemical potential 
$\mu=-0.016t$ (see text). 
Solid line: the $T$-dependence of the radius of the 
gapped region $R(T)\propto (T^{*}-T)$, the incommensuration $\delta 
Q=0.12\pi$. Dashed line: $R(T)\propto (T^{*}-T)$, $\delta Q=0$.
Dotted line: $R(T)\propto (T^{*}-T)^{1/2}$,  $\delta Q=0.12\pi$.
Dashed-Dotted line: $(T^{*}-T)$, $\delta Q=0.12\pi$.
\end{document}